# Evolution of overlapping resonances along an isoelectronic sequence

W.-C. Chu[1], H.-L. Zhou[1], A. Hibbert[2], and S. T. Manson[1]

[1]*Department of Physics and Astronomy, Georgia State University, Atlanta, Georgia 30303, USA*

[2]*School of Mathematics and Physics, Queen's University of Belfast, Belfast BT7 1NN, United Kingdom*

## Abstract

Near-threshold resonances have been studied for Be-like ions with a focus on overlapping resonances among Rydberg series converging to different thresholds. The behavior of the overlapping as a function of *Z* and the approach to the limit of infinite *Z* are investigated. The $4s4p$ resonance is shown and discussed in detail as an example.

## I. INTRODUCTION

In the experimental or theoretical study of atomic or ionic photoionization, it is found that the cross sections are often dominated by resonances in many photon energy regions, especially the near-threshold ones. The resonances are extremely informative of the dynamics of the systems, but in many situations the information is hard to extract due to the strong perturbation engendered by nearby resonances, insufficient energy mesh points, or other issues. The overlapping of the resonance series converging to different thresholds complicates the analysis of resonances. Since the neutral atoms are more intensely studied, the aim of this investigation is to inquire how perturbations affect resonances in ions.

In particular, we study the perturbation of the resonances for Be-like ions, and focus on the evolution along the isoelectronic sequence. This sequence is chosen because 4-electron systems are simple enough so that the calculations are highly accurate, and recent experiments [1-7] offer a large amount of data. Good agreement has been found between the measurements and the calculations, thereby ensuring that the present study is quantitatively accurate. The total cross sections of the 14 ions, Be, $B^{+}$, $C^{+2}$, $N^{+3}$, $O^{+4}$, $Ne^{+6}$, $Mg^{+8}$, $Si^{+10}$, $S^{+12}$, $Ar^{+14}$, $Ca^{+16}$, $Ti^{+18}$, $Cr^{+20}$, and $Fe^{+22}$, were calculated using the relativistic Breit-Pauli *R*-matrix method [8] and were reported earlier [9]. To look closely at the overlapping resonance profiles, the cross section for each ion was calculated on an evenly distributed set of $10^5$ photon energy points between the 2*s* ionization threshold and the 4*f* threshold. Considering the channel interactions and the relativistic effects, the overlapping of resonances along the sequence are quite complicated, especially in the near-threshold regions where resonances are tightly crowded, but the analysis is aided by knowing the asymptotic behavior in the limit of infinite *Z*. At the $Z \rightarrow \infty$ limit, we can calculate exactly where the resonances converging to higher thresholds will appear as interlopers and perturbers among the resonances converging to lower thresholds. Since it is expected that the general appearance of the cross section changes with *Z* smoothly along an isoelectronic sequence, this asymptotic behavior allows us to trace the positions of specific resonances along the sequence in the energy regions where a strong perturbation may blur or distort the resonance profiles. The 4*s*4*p* resonance is used in this paper as an example to show how the overlapping evolves with *Z*.

Sec. II briefly describes the theory and the calculation method that is employed in this theoretical work. In Sec. III, a derivation of the overlapping positions between different series at infinite $Z$ is presented, along with the example of the evolution of the $4s4p$ resonance. The conclusions are given in Sec. IV.

## II. THEORY

In this study the photoionization of the ground ${}^1S_0^{\mathrm{e}}$ state of the Be-like systems are calculated, i.e.,

$$1s^2 2s^2 \left({}^1S_0^{\mathrm{e}}\right) + h\nu \rightarrow \left[1s^2 nl + e^-(kl')\right]\left({}^1P_1^{\mathrm{o}}\right). \tag{1}$$

The computational package employed is the RMATRX1 program [8], based on the Breit-Pauli $R$-matrix method [10]. The essential idea of $R$-matrix theory is to divide configuration space into internal and external regions by a spherical shell of radius $a$ centered at the nucleus, and connect the wave functions between the regions by means of the $R$-matrix. In the internal region, since all the $N+1$ electrons are relatively close to each other, they are considered indistinguishable and all the exchange terms between them are included in the calculation of the wave function. In the external region, setting $a$ large enough, all the discrete wave functions are assumed to be zero, and the whole system is treated as a two-body system, which consists of the continuum photoelectron and the $N$-electron "target". In this way, the total wave function at any given energy $E$ is obtained. Using the initial and final state wave functions obtained, the photoionization cross section is then calculated in the electric dipole approximation.

To construct accurate wave functions for the $N$-electron target states, 9 configurations, $1s^2 2s$, $1s^2 2p$, $1s^2 3s$, $1s^2 3p$, $1s^2 3d$, $1s^2 4s$, $1s^2 4p$, $1s^2 4d$, and $1s^2 4f$, are included as the basis to solve the Schrödinger equation. The associated 10 orbitals are optimized using the CIV3 program [11]. The target state wave functions are rather accurate; quite good agreement between the target state energies and the NIST values was reported in Ref. [9]. Then the $(N+1)$-electron total wave functions are constructed by adding one more electron to the $N$-electron wave functions. In the internal region, antisymmetry is considered and all exchange terms are included; in the external region, exchange between the photoelectron and the target is omitted, and the boundary conditions at $r \rightarrow \infty$ are applied. The connection between the regions is made by requiring that the total $(N+1)$-electron wave function and its first derivative are continuous at $r=a$. The energies of the $(N+1)$-electron discrete states are also shown in Ref. [9], which are in good agreement with experiments, thereby confirming the quality of the wave functions.

In Eq. (1), the states are specified by the $LSJ\pi$ symmetries. However, in the relativistic Breit-Pauli *R*-matrix calculations, the states are specified by only the $J\pi$ terms, where each $J\pi$ term has the contributions from all the possible $LS$ terms. Thus, the ground state transition is actually defined as $0^{\mathrm{e}} \to 1^{\mathrm{o}}$.

## III. RESULTS AND DISCUSSION

It is of interest to study which resonances in a series that converges to a given threshold lie below the lower threshold, thereby perturbing the resonance series converging to the lower threshold. Further, if some of the resonances are embedded in a series converging to a lower threshold, it is of interest to determine their positions among the resonances of the lower series. The analysis starts with the asymptotic region, $Z\to\infty$, where some definite predictions can be made. As $Z\to\infty$, the electron-electron interactions in the Hamiltonian of an atomic system become infinitesimal compared with the nuclear potential terms, and the system becomes hydrogenic [12]. The energy levels of each electron in a hydrogenic atomic system of nuclear charge $Z$ is given by

$$E_n = -\frac{Z^2}{n^2}, \tag{2}$$

where $n$ is an integer. In this hydrogenic system, the energy of an autoionizing two-electron excitation is given by

$$E_{n,n'} = -\frac{Z^2}{n^2} - \frac{Z^2}{n'^2}, \tag{3}$$

where $n$ and $n'$ are both integers, i.e., eq. (3) represents the $nln'l'$ resonance energy. However, these energies are independent of the angular momenta. Thus, the condition that the $nln'l'$ resonance lies below the $n-1$ threshold is

$$-\frac{Z^2}{(n-1)^2} > -\frac{Z^2}{n^2} - \frac{Z^2}{n'^2} \tag{4}$$

or equivalently,

$$\frac{1}{(n-1)^2} < \frac{1}{n^2} + \frac{1}{n'^2}. \qquad (5)$$

Applying this simple relation to the resonances in Be-like systems, it is found that for the $n=2$ thresholds, no resonances converging to $n>2$ thresholds lie below the $n=2$ thresholds. However, for $n=3$, the $4l4l'$ resonances lie below this threshold and perturb the $3ln'l'$ resonance series. For the $n=4$ thresholds, it is found that the $5l5l'$ and the $5l6l'$ resonances lie below the $n=4$ thresholds. This analysis can be carried to higher thresholds as well, and the results are summarized in Table I. From this Table it is clear that, with increasing *n*, more and more resonances converging to the next higher threshold lie below the lower threshold. While these conclusions are exact in the $Z\rightarrow\infty$ hydrogenic limit, they are pretty close to what happens for finite *Z* as well, particularly for highly charged ions.

Using these same ideas, the actual positions of these interloping resonances compared to the lower resonance series can be ascertained in the $Z\rightarrow\infty$ hydrogenic limit. For example, the energy of the $4l4l'$ resonances, designated $E_{4,4}$, lies between $E_{3,8}$ and $E_{3,9}$, i.e., the $4l4l'$ resonances lies between the $3l''8l'''$ resonances and the $3l''9l'''$ resonances. Similarly, the $5l5l'$ resonances is situated between $4l''7l'''$ and $4l''8l'''$, $6l6l'$ between $5l''8l'''$ and $5l''9l'''$, etc.

The first few resonances in a Rydberg series can be positioned below a lower ionization threshold and, consequently, are strongly perturbed by the resonances converging to the lower threshold. These strongly perturbed resonances are very difficult to characterize and to identify because their profiles are dramatically changed by all the resonances of the lower series. As an example, Fig. 1 shows the evolution of the 4*s*4*p* resonance profile for ground state photoionization of six of the lower members of the Be isoelectronic series, which is an excellent example of this type of strong perturbation. The top two panels present the cross sections in the vicinity of the 4*s*4*p* resonance for Be and $B^+$. On the lower photon energy side of the 4*s*4*p* resonance are the 3*d*, the 3*p*, and the 3*s* thresholds (they are not shown for Be because they are much lower than the energy range shown), from right to left, which are characterized by the convergence of the 3*dnl*, the 3*pnl*, and the 3*snl* resonances. Because these thresholds are all below the 4*s*4*p* resonance for Be and $B^+$, the interaction of the lower series of resonances with the 4*s*4*p* resonance is quite small owing to the large energy denominator in the mixing coefficient. Thus the perturbation to 4*s*4*p* resonances is quite small, and the resonance profile is clean. The third panel from top is the cross section for $C^{+2}$. In this case the 4*s*4*p* resonance lies between the 3*p* and the 3*d* thresholds, so that a number of the 3*dnl* resonances, which are nearly degenerate with the 4*s*4*p* resonance, perturb the 4*s*4*p* resonance and build many narrow "spikes" on its profile. The 4*s*4*p* profile is still clearly recognizable since it is much wider than all the narrow resonances converging to the 3*d* threshold, but the characterization is harder and definitely less clear than that in the case of Be and $B^{+}$. The fourth panel from top is for $N^{+3}$. Here the 4*s*4*p* resonance lies between the 3*s* and the 3*p* thresholds, and its profile is much messier, where it overlaps not only 3*dnl* but also 3*pnl* resonances. The narrow resonances sort of cut the 4*s*4*p* profile to short pieces, and the 4*s*4*p* profile becomes the "background" or modulation of those narrow resonances. In other words, what is happening is that the wave functions of the resonances in the region are strongly mixed between 3*dnl* (or 3*pnl*) and 4*s*4*p* so there is no longer a pure 4*s*4*p* resonance. We can estimate where this mixed 4*s*4*p* resonance exists very roughly, but the complete characterization is extremely difficult. The fifth panel from top shows the cross section for $O^{+4}$. The 4*s*4*p* resonance in this case lies between the 3*s* and the 3*p* thresholds; its profile is quite flat, and its width is difficult to pick out among the surrounding (perturbing) resonances. The bottom panel is the cross section for $Ne^{+6}$, and here the 4*s*4*p* resonance is situated below the 3*s* threshold, so it overlaps all of the resonances series converging to the 3*l* thresholds. It is seen that it is almost completely hidden among the $3lnl'$ resonances. Thus, although the 4*s*4*p* resonance does move below the $n=3$ threshold with increasing *Z*, as expecting from the analysis of the asymptotic $Z\to\infty$ case above, as the resonance moves down below lower thresholds, it also gets very mixed until its identity is virtually gone.

Nevertheless, the quantum defect $\mu$ (and the associated effective quantum number $\nu$) of a resonance changes with *n* and with *Z* smoothly, which provides us a way to trace the resonance positions as a function of *Z*. As an example, the cross section in the vicinity of the 4*s*4*p* resonance is displayed in Fig. 2 as a function of $\nu$ with respect to the 4*s* threshold ($\nu_{4s}$) for the same six ions as in Fig. 1 to demonstrate the smooth trajectory of the 4*s*4*p* resonance along the sequence. The quantum defect $\mu$ for the 4*s*4*p* resonance goes to zero as $Z\to\infty$ (as all quantum defects must do), which means the associated $\nu$ converges to 4 asymptotically. By knowing the $\nu$'s for the lowest-$Z$ ions and by the convergence of $\nu$ toward $\nu = 4$ as $Z \to \infty$, the 4*s*4*p* resonance positions for all ions in the sequence can be accurately estimated since the shifts along the sequence are monotonic and smooth with respect to $\nu$. In addition, the 4*s*4*p* resonance width (in $\nu$) remains almost constant with increasing $Z$.

## IV. CONCLUDING REMARKS

Using hydrogenic theory, resonance positions with respect to lower thresholds have been analytically calculated for $Z \to \infty$, and predictions were made as to which resonances overlap with lower series along with the positions of these overlapping resonances with respect to the resonances of lower series.

In addition, using our earlier calculations of the photoionization cross sections of the ${}^1S_0^{\mathrm{e}}$ ground state of Be-like ions [9], the position of the 4*s*4*p* resonance along the sequence is analyzed as a function of both photon energy and effective quantum number $\nu$. The quantum defect of the resonance is found to decrease smoothly and monotonically with increasing $Z$, and the position of the resonance is seen to be converging to $\nu = 4$ as predicted by hydrogenic theory. At the higher $Z$ values, all resonances in all isoelectronic sequences will converge monotonically to zero quantum defect, but near the neutral end of the sequence, where the electron-electron correlation is important and the wave functions are complicated, it is possible for quantum defect to change non-monotonically with $Z$, unlike the present case.

The 4*s*4*p* is an isolated resonance in the photoionization of Be and $B^+$, but as $Z$ increases, it begins to overlap lower Rydberg series converging to the $n = 3$ thresholds. This overlap becomes greater with increasing $Z$; consequently, the resonance becomes more and more strongly perturbed, and the resonance profile becomes more and more distorted by the interactions with the lower series and less recognizable as a resonance.

While the analysis of a particular resonance arising from the photoionization of the Be-like sequence has been detailed in this study, the methodology employed should be applicable to other resonances in this and other isoelectronic sequences as well. On the experimental side, since photoionization cross section measurements of many of the ions are now possible with excellent resolution, it would be of great interest to see experimental investigation of the 4*s*4*p* resonance along the isoelectronic sequence.

## Acknowledgments

This work was supported by U.S. Department of Energy, Division of Chemical Sciences, and NSF. All calculations were performed on National Energy Research Scientific Computing Center (NERSC) computers.

**Figure Captions**

1. (color online) Calculated photoionization cross sections near the 4*s*4*p* resonances for the ground states of Be, $B^{+}$, $C^{+2}$, $N^{+3}$, $O^{+4}$, and $Ne^{+6}$. The 4*s*4*p* resonance profiles are encircled. The vertical dashed lines indicate the various $n$ = 3 thresholds. The resonance starts overlapping the $n = 3$ series in $C^{+2}$, and is more and more perturbed as $Z$ increases.
2. (color online) As of Fig. 1, but plotted against effective quantum number $\nu$ with respect to 4*s* threshold. Note that the cross sections are on log scale. The long vertical dashed line indicates $\nu = 4$, and the short lines are for the various $n$ = 3 thresholds in each case. The 4*s*4*p* resonance positions are seen to converge to $\nu = 4$ asymptotically.

Table I. Resonances lying below the next lower threshold in hydrogenic systems.

| $n$ | Resonances Below the Next Lower Threshold |
|---|---|
| 3 | $4l4l'$ |
| 4 | $5l5l', 5l6l'$ |
| 5 | $6l6l'$ - $6l9l'$ |
| 6 | $7l7l'$ - $7l11l'$ |
| 7 | $8l8l'$ - $8l14l'$ |
| 8 | $9l9l'$ - $9l17l'$ |
| 9 | $10l10l'$ - $10l19l'$ |
| 10 | $11l11l'$ - $11l24l'$ |

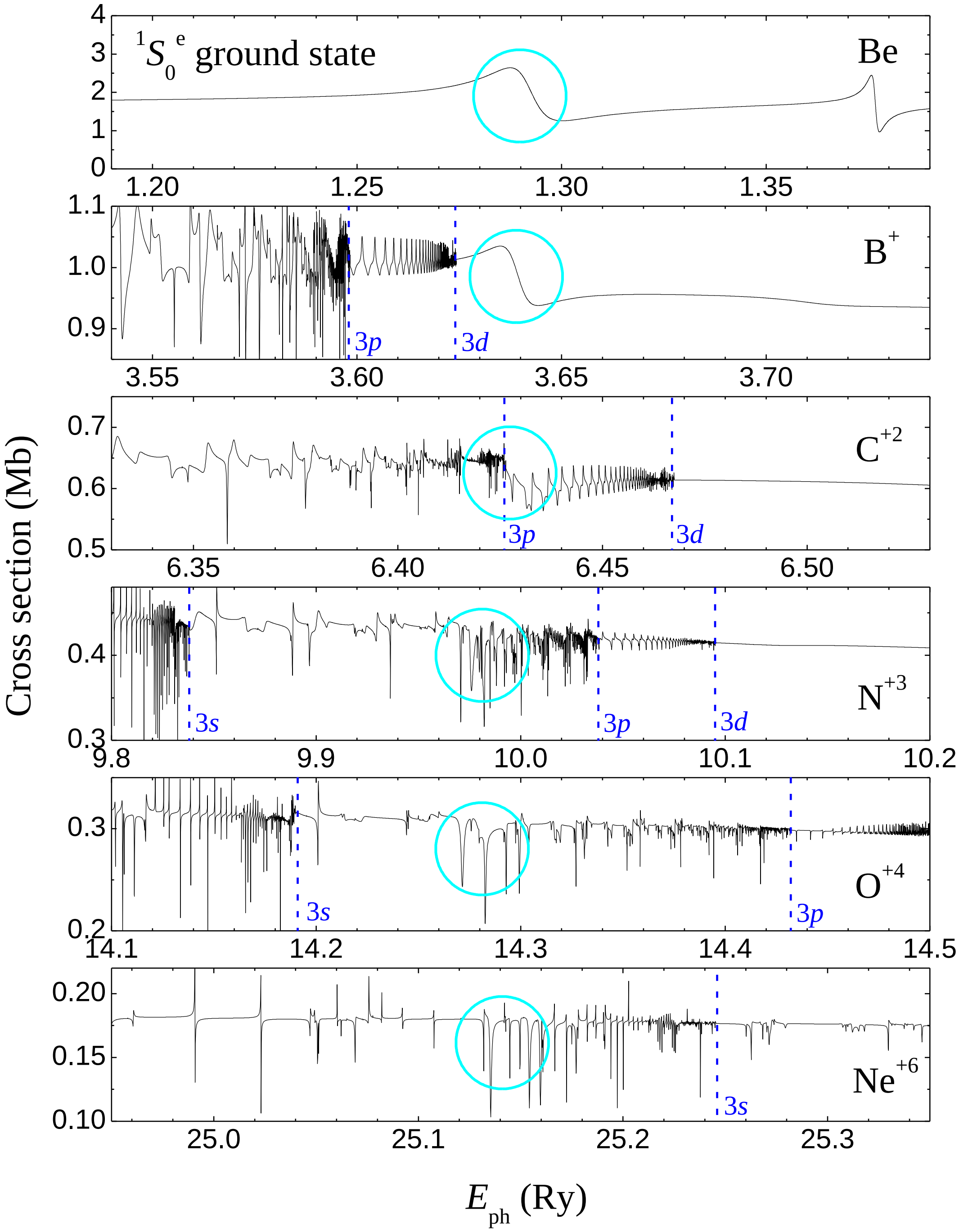

Fig. 1

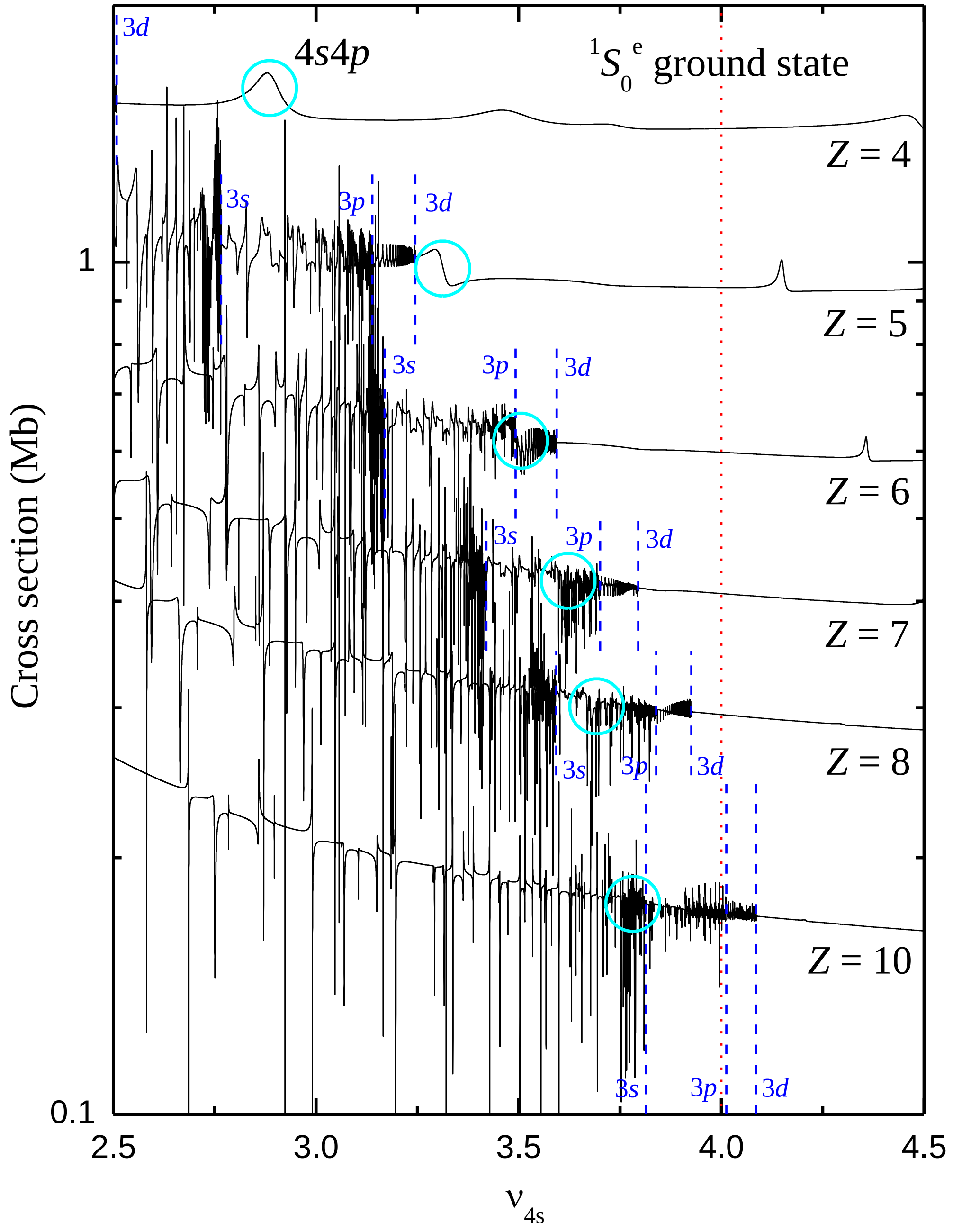

Fig. 2